\documentclass[twocolumn,prl,superscriptaddress,floatfix,showpacs]{revtex4-1}
\usepackage[utf8]{inputenc}
\usepackage{amsmath}
\usepackage{amssymb}
\usepackage{graphicx}

\usepackage{graphics}
\usepackage{epsfig}

\newcommand{\be}{\begin{equation}} \newcommand{\ee}{\end{equation}}
\newcommand{\bea}{\begin{eqnarray}} \newcommand{\eea}{\end{eqnarray}}

\begin{document}

\title{How fast does a random walk cover a torus?}

\author{Peter Grassberger}

\affiliation{JSC, FZ J\"ulich, D-52425 J\"ulich, Germany}

\date{\today}
\begin{abstract}
We present high statistics simulation data for the average time $\langle T_{\rm cover}(L)\rangle$ 
that a random walk needs to cover completely a 2-dimensional torus of size $L\times L$. They 
confirm the mathematical prediction that $\langle T_{\rm cover}(L)\rangle \sim (L \ln L)^2$ for 
large $L$, but the prefactor {\it seems} to deviate significantly from the supposedly exact 
result $4/\pi$ derived by A. Dembo {\it et al.}, Ann. Math. {\bf 160}, 433 (2004), if the most 
straightforward extrapolation is used. On the other hand, we find that this scaling does hold for 
the time $ T_{\rm N(t)=1}(L)$ at which the average number of yet unvisited sites is 1, as also 
predicted previously. This might suggest (wrongly)
that $\langle T_{\rm cover}(L)\rangle$ and $T_{\rm N(t)=1}(L)$ scale differently,
although the distribution of rescaled cover times becomes sharp in the limit $L\to\infty$. But our
results can be reconciled with those of Dembo {\it et al.} by a very slow and 
{\it non-monotonic} convergence of $\langle T_{\rm cover}(L)\rangle/(L \ln L)^2$, as had been
indeed proven by Belius {\it et al.} [Prob. Theory \& Related Fields {\bf 167}, 1 (2014)] for 
Brownian walks, and was conjectured by them to hold also for lattice walks.
\end{abstract}
\maketitle

%\section{Introduction}

The problem of how fast a random walk covers a 2-dimensional torus was introduced in the mathematical
literature by Wilf \cite{Wilf}, who called it the ``white screen problem". But it is also of considerable
interest for other sciences, as it relates e.g. to how fast a grazing animal can collect as much food
as possible \cite{Viswanathan,Santos,Benichou}, or how fast information can be spread on or collected
from a network (such as a mobile ad hoc network) whose topology is not known \cite{Mian,Li,Avin}. For that
reason, it has also been discussed extensively in the statistical physics literature
\cite{Nemiro,Brummelhuis-a,Brummelhuis-b,Coutinho,Mendonca,Chupeau}.

Let us denote by 
$\langle T_{\rm cover}(L)\rangle$ the average time needed to cover a torus of $L\times L$ sites 
completely, and by $ T_{\rm N(t)=1}(L)$ the time at which the average number
of yet uncovered sites is 1. Naively one would expect that both diverge in the same way with $L$,
at least if the distribution of cover times is not too broad.

Aldous \cite{Aldous-a,Aldous-b} proved that 
\be
    \langle T_{\rm cover}(L)\rangle \lesssim \frac{4}{\pi} L^2 \ln^2 L,
\ee
and proved that the re-scaled time, $T_{\rm cover}(L)/(L\ln L)^2$, is indeed $\delta$-distributed 
in the limit $L\to\infty$. He furthermore conjectured that Eq.(1) becomes sharp in this limit. 

This conjecture was supported by heuristic arguments in \cite{Brummelhuis-a,Brummelhuis-b}, where the 
main quantity of interest was not $\langle T_{\rm cover}(L)\rangle$ but $T_{\rm N(t)=1}(L)$. 
These authors argued convincingly that 
\be
   T_{\rm N(t)=1}(L) /(L\ln L)^2 \to \frac{4}{\pi}  \quad {\rm for} \;\; L \to\infty,
              \label{brummel}
\ee
and then conjectured that the same is true also for the cover times, because mean cover times
and times at which the average number of uncovered sites is 1 should scale in the same way.

The story was seemingly closed when Dembo {\it et al.} \cite{Dembo} proved rigorously that 
\be
    \lim_{L\to\infty}\frac{T_{\rm cover}(L)}{(L \ln L)^2} = \frac{4}{\pi} \;\;\; 
       in \;\;probability,  \label{dembo}
\ee
i.e. Aldous' inequality Eq.(1) is saturated and the limit distribution is indeed sharp.

When I re-considered this problem, I was primarily interested in the way how ``true self avoiding"
walks (or ``self-repelling walks") \cite{Amit} cover the torus or any other finite lattice
\cite{Freund,Avin}, and wanted just to document the dramatic difference between self-repelling and
ordinary random walks. However, soon after I started to simulate ordinary random walks on the 2-torus, 
it became clear that the data agreed with Eqs.(1) and (\ref{brummel}), but not easily with 
Eq.(\ref{dembo}).

The results presented in the following come from simulations that altogether took about 1 year of CPU 
time on modern workstations. Lattice sizes ranged from $L=16$ to $L=65536$ in steps of powers of 2.
The number of walks simulated varied between $\approx 4\times 10^7$ for $L=16$ and $1350$ for $L=65536$.
For easier coding and faster codes, boundary conditions (b.c.) were not strictly periodic but helical 
\cite{New-Bark}. For large $L$ the difference is negligible. In particular, also for helical b.c.
the lattice is a torus, and the difference with periodic b.c. is just that one of the coordinate
axes is slightly tilted. We verified that the results obtained with periodic b.c. were identical 
within statistical errors for $L\geq 16$. We also tested two different random number generators 
(Ziff's four-tap 
generator \cite{Ziff} and the UNIX generator rand48), again with no significant differences.

\begin{figure}
\begin{centering}
\includegraphics[scale=0.30]{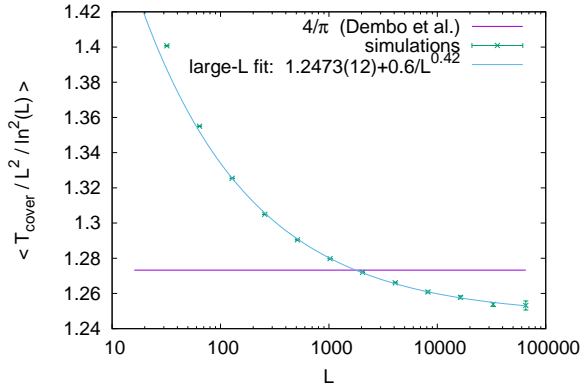}
\par\end{centering}
\caption{\label{fig1} (color online)  Log-linear plot of average cover time for 2-toruses of size 
  $L\times L$, plotted against $L$. Whenever no error bars are visible on the data points, they are
  smaller than the line thickness. The fit is for all data with $L\geq 128$. The error of the 
  leading term is supposed to take into account the possibility of further corrections to scaling
  that would, however, leave the cover times monotonically decreasing with $L$.}
\end{figure}

Results for $\langle T_{\rm cover} / (L \ln L)^2 \rangle$ against $L$ are shown in Fig.~1. Whenever 
error bars are not visible on the data points, they are smaller than the line thickness. Also 
shown is the prediction of Dembo {\it et al.} \cite{Dembo} (horizontal line) and a fit for large $L$.
This fit is a least square fit (with all three constants fitted) to all data with $L\geq 128$, but 
the quoted error in the first term is much bigger that the purely statistical error, in order to 
include plausible further correction terms -- where we assume that ``plausible" correction terms do 
not ruin the monotonicity. Our first conclusion is thus that 
\be
    \lim_{L\to\infty}\frac{\langle T_{\rm cover}(L) \rangle}{(L \ln L)^2} = 1.2473 \pm 0.0012.
\ee

The right hand side disagrees with the supposedly exact value $4/\pi = 1.2732 \ldots$ by about 22
standard deviations (similar results have been obtained in \cite{Mendonca}, albeit with less 
statistics). This discrepancy can hardly be blamed on statistical fluctuations (the 
likelihood being about $10^{-100}$). It cannot be blamed on the used random number generators,
both of which have been proven to be reliable even in problems involving much higher statistics.
In view of the extreme simplicity of the code (about one page), also a programming error is very 
unlikely. 

\begin{figure}
\begin{centering}
\includegraphics[scale=0.30]{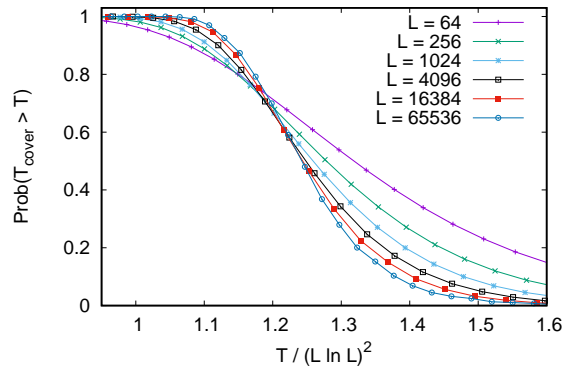}
\par\end{centering}
\caption{\label{fig2} (color online)  Cumulative distributions ${\rm Prob}(T_{\rm cover} >T)$ for 6 values of $L$.}
\end{figure}

\begin{figure}
\begin{centering}
\includegraphics[scale=0.30]{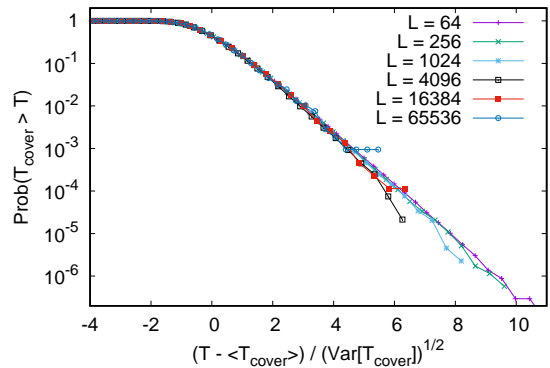}
\par\end{centering}
\caption{\label{fig3} (color online)  Same data as in Fig.~2, but (i) plotted on a logarithmic y-scale;
  and (ii) plotted against $(T - \langle T_{\rm cover} \rangle)/\{{\rm Var}[T_{\rm cover}]\}^{1/2}$.}
\end{figure}

A next problem that could cause a wrong asymptotic estimate could be a very skewed and broad 
distribution of cover times. But the distribution of normalized cover times is expected \cite{Belius-a} 
to be a (randomly shifted) Gumbel distribution in the limit $L\to\infty$. This gives a roughly 
exponential tail, which could not significantly bias any estimates of average cover times.

In any case, in Figs.~2 and ~3 we show such distributions. They seem to be indeed exponentially 
cut off at large times, and definitely do not suggest that estimates of the averages could be influenced 
significantly by large $T$ tails.

\begin{figure}
\begin{centering}
\includegraphics[scale=0.30]{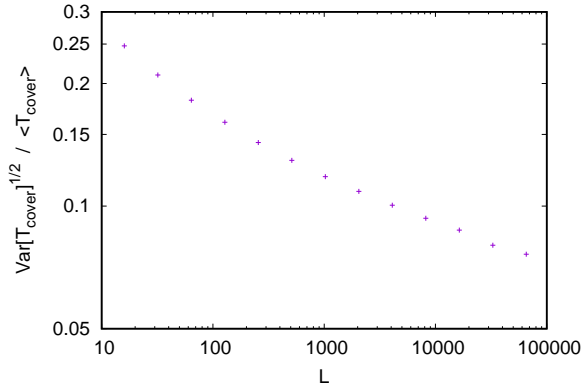}
\par\end{centering}
\caption{\label{fig4} (color online)  Relative fluctuations of cover times, plotted against $L$.}
\end{figure}

To add to the last point, we show in Fig.~4 our estimates of the relative fluctuations of 
$T_{\rm cover}$, defined as $\{{\rm Var}[T_{\rm cover}]\}^{1/2} / \langle T_{\rm cover} \rangle$.
We see that they decrease with $L$ as predicted by Aldous, although our data are not precise
enough to distinguish between a power-law decay with a very small exponent ($\approx 0.11$)
and a logarithmic behavior.

\begin{figure}
\begin{centering}
\includegraphics[scale=0.30]{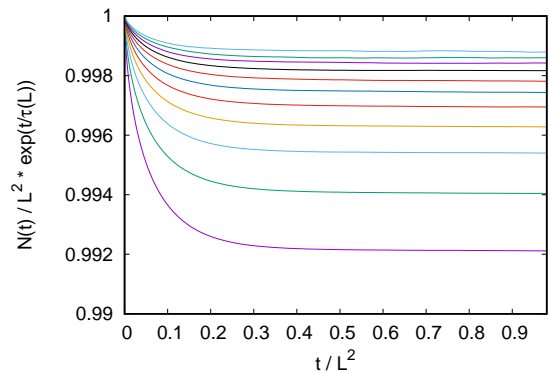}
\par\end{centering}
\caption{\label{fig5} (color online)  The average number $N(t)$ of uncovered sites at time, 
  plotted for different values of $L$ in the regime $t < L^2$. For clarity, we show on the y-axis
  not $N(t)$ itself but $N(t)/L^2 \exp(t/ \tau(L))$, where $\tau(L)$ is the numerically found inverse
  decay rate of $N(t)$ for $t \gg L^2$. The uppermost curve is for $L=65536$, the lowest is for $L=64$.}
\end{figure}

\begin{figure}
\begin{centering}
\includegraphics[scale=0.30]{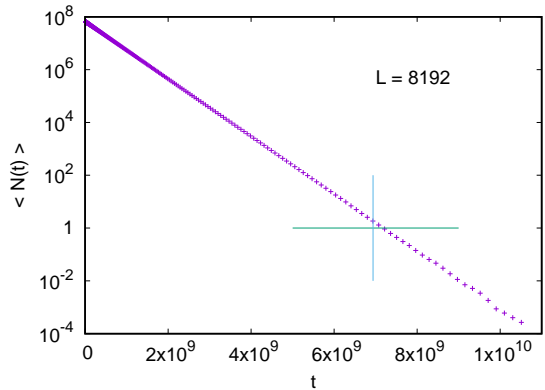}
\par\end{centering}
\caption{\label{fig6} (color online)  The average number $N(t)$ of uncovered sites at time, plotted
  against $t$, for $L=8192$. The horizontal and vertical straight lines indicate the values 
  $N(t)=1$ and $t = \langle T_{\rm cover} \rangle$.}
\end{figure}

To shed more light on this problem, we considered next the average number $N(t)$ of uncovered sites
at time $t$. For $t \ll L^2$, the number of {\it covered} sites is independent of $L$, and given
asymptotically by \cite{Dvoretsky}
\be
   s(t) \equiv L^2 - N(t) = \frac{\pi t}{\ln t} \left[ 1 + O(\frac{\ln \ln t}{\ln t})\right].
\ee
The finiteness of the lattice becomes relevant for $t \approx L^2$, and for $t \gg L^2$ the decay
of $N(t))$ is a pure exponential \cite{Brummelhuis-a}. The cross-over between these two regimes is 
shown in Fig.~5. There we show on the y-axis not $N(t)/L^2$ itself, but we multiplied it with 
$\exp(t/ \tau(L))$, where the characteristic time $\tau(L))$ (the inverse decay rate) was 
estimated from fits in the regime $L^2 < t < (L \ln L)^2$. The quality of the exponential decay 
in this regime is illustrated in Fig.~6 for $L=8192$ (but similarly nice exponentials were also 
found for all other lattice sizes). In Fig.~6 we plotted $N(t)$ itself, and we verified that 
the exponential decay continued also for $t \gg (L \ln L)^2$, although statistical errors increase
rapidly for large $t$.

\begin{figure}
\begin{centering}
\includegraphics[scale=0.30]{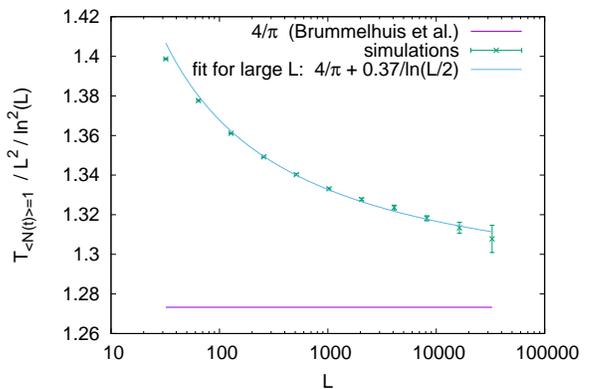}
\par\end{centering}
\vglue -5mm
\caption{\label{fig7} (color online) Direct estimates of $T_{\rm N(t)=1} / (L\ln L)^2$ plotted against
  $L$ on a log-linear plot. The fit just demonstrates that the data are compatible with Eq.(\ref{brummel}).}
\end{figure}

This purely exponential decay can be used to determine $\tau(L)$ either by a fit in the regime 
$L^2 < t < (L \ln L)^2$ or by just finding the value of $t$ where $N(t)=1$. In the second 
method we of course have to take into account that the exponential decay holds only for 
$t > L^2$, but this correction becomes negligible for $L\to\infty$, i.e. 
\be
   T_{\rm N(t)=1} = 2\tau(L) \ln L \times [1 + O(1/\ln^2L)].
\ee
Direct numerical estimates of $T_{\rm N(t)=1} / (L\ln L)^2$ are shown in Fig.~7. We see a much slower 
(probably logarithmic) convergence than for average cover times, but the data are completely 
compatible with Eq.(\ref{brummel}).

\begin{figure}
\begin{centering}
\includegraphics[scale=0.30]{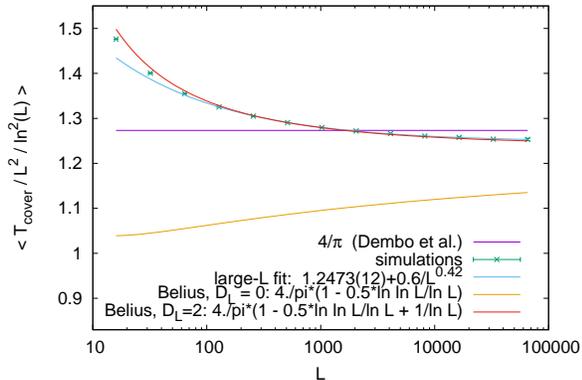}
\par\end{centering}
\vglue -5mm
\caption{\label{fig8} (color online) Same data as in Fig.~1, but with two additional analytic curves. Both
  represent Eq.(\ref{Belius}), one with $D=0$ and the other with $D=2$.}
\end{figure}

A last reason for a wrong asymptotic estimate would be a very slow (and non-monotonic!) convergence 
with $L$. We found no indication for this in our data, but it is conjectured in \cite{Belius-b} that the 
behavior for walks on the square lattice is as for off-lattice Brownian walks, 
which would suggest \cite{Belius-b}
\be
    \frac{\pi\langle T_{\rm cover}(L)\rangle}{4(L \ln L)^2} = 1 -\frac{1}{2}\ln \ln L/\ln L + D /\ln L +
      o(1/\ln L)   \label{Belius}
\ee
with an unknown constant $D$ (indeed, the conjecture in \cite{Belius-b} for lattice walks was slightly weaker). 
In Fig.~8 we show the data shown already in Fig.~1 together with two analytic curves representing 
Eq.(\ref{Belius}): One with $D=0$, and the other with $D=2$. We see that the latter gives a very good fit,
from which we conclude that the mathematical predictions are presumably all correct, and $D=2.02(2)$. We 
should warn, however, that we could also give decent fits with different coefficients of the 
$\ln \ln L/\ln L$ term (and, of course, different $D$).

\begin{figure}
\begin{centering}
\includegraphics[scale=0.30]{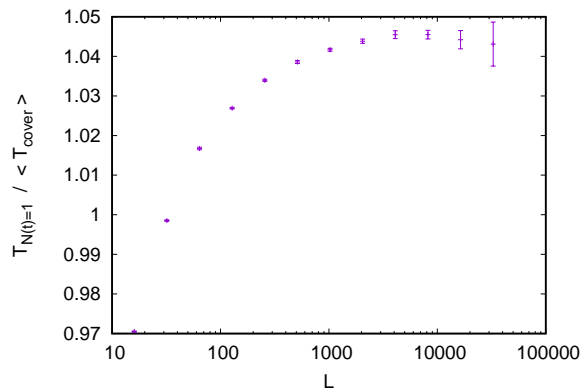}
\par\end{centering}
\vglue -5mm
\caption{\label{fig9} (color online) Ratios $T_{\rm N(t)=1} / \langle T_{\rm cover}\rangle$ plotted against
  $L$ on a log-linear plot.}
\end{figure}

Finally, we show in Fig.~9 the ratios $T_{\rm N(t)=1} / \langle T_{\rm cover}\rangle$.
For very small $L$ they are $<1$, because the large-$T$ tails contribute more to $\langle T_{\rm cover}\rangle$
than to $T_{\rm N(t)=1}$. For larger $L$ this effect is outweighed by the fact that
$N(\langle T_{\rm cover}\rangle) >1$ because walks that do not yet cover at $t=\langle T_{\rm cover}\rangle$
might have $\gg 1$ uncovered sites. Finally, at very large $L$, the ratio seems to decrease again,
although this is not significant in view of the large error bars. Yet it suggests that the ratio converges
to 1 for $L\to\infty$, which would completely reconcile our data with the mathematical proofs. This is 
supported by the fact (O. Zeitouni, private communication) that the $\ln \ln L/\ln L$ term is absent in 
$T_{\rm N(t)=1}$.

In summary, our numerical data suggest {\it at face value} that $T_{\rm N(t)=1}$ and $\langle T_{\rm cover}\rangle$
do not scale in the same way with $L$, in contrast to rigorous proofs. But they can be reconciled
with the proofs, if the (predicted) corrections to scaling are taken into account. As a result, the convergence
towards the asymptotic behavior should be extremely slow (and non-monotonic!). Thus, without knowing the 
subleading terms, attempts to verify the leading behavior numerically would be futile.

The present paper can be seen as a warning that supposedly rigorous proofs {\it can} be wrong (and should
thus be checked numerically), but more so as a warning that extrapolations of numerical data can be 
very subtle and misleading, even if they {\it look} completely benign and harmless. The vast number of 
wrong critical exponent estimates found in the literature bears ample witness to that. Combining rigorous
mathematics and numerics can be useful if, as in the present case, the mathematics exclude too naive 
parametrizations, and the numerics can suggest the value(s)
of constants that remain undetermined by the mathematical arguments.

I thank Pradep K. Mohanty, Bob Ziff, and Ofer Zeitouni for carefully reading the manuscript, and to Ricardo 
Mendon\c{c}a for pointing out Ref.~\cite{Mendonca}. To all of them and also to David Belius, I am indebted for 
extremely helpful discussions.


\begin{thebibliography}{10}
\bibitem{Wilf} H.S. Wilf, Amer. Math. Monthly {\bf 96}, 704 (1989).
\bibitem{Viswanathan} G.M. Viswanathan, S.V. Buldyrev, S. Havlin, M.G.E. Da Luz, E.P. Raposo, and 
    H.E. Stanley, Nature {\bf 401}, 911 (1999).
\bibitem{Santos} M.C. Santos, G.M. Viswanathan, E.P. Raposo, and M.G.E. da Luz, Phys. Rev. E {\bf 72}, 046143
    (2005). 
\bibitem{Benichou} O. B\'enichou, C. Loverdo, M. Moreau, and R. Voituriez, Rev. Mod. Phys. {\bf 83}, 81 (2011).
\bibitem{Mian} A.N. Mian, R. Beraldi, and R. Baldoni, 2010 IEEE 7th Int'l Conf. on Mobile Adhoc and Sensor 
    Systems, p. 146 (2010).
\bibitem{Li} K. Li, Int'l J. of Foundations of Computer Science {\bf 23}, 779 (2012).
\bibitem{Avin} C. Avin and B. Krishnamachari, Computer Networks {\bf 52}, 44 (2008).
\bibitem{Nemiro} A.M. Nemirovsky, H.O. M\'artin, and M.D. Coutinho-Filho, Phys. Rev. {\bf A 41}, 761 (1990).
\bibitem{Brummelhuis-a} M.J.A.M. Brummelhuis and H.J. Hilhorst, Physica A {\bf 176}, 387 (1991).
\bibitem{Brummelhuis-b} M.J.A.M. Brummelhuis and H.J. Hilhorst, Physica A {\bf 185}, 35 (1992).
\bibitem{Coutinho}
\bibitem{Mendonca} J.R.G. Mendon\c{c}a, Phys. Rev. E {\bf 84}, 022103 (2011).
\bibitem{Chupeau} M. Chupeau, O. B\'enichou, and R. Voituriez, Nature Physics {\bf 11}, 844 (2015).
\bibitem{Aldous-a} D. Aldous, {\it Probability approximations via the Poisson clumping heuristic},
   Applied Math. Sci. {\bf 77} (Springer, New York 1989).
\bibitem{Aldous-b} D. Aldous, {\it Threshold limits for cover times}, J. Theor. Probab. {\bf 4}, 
   197 (1991).
\bibitem{Dembo} A. Dembo, Y. Peres, J. Rosen, and O. Zeitouni, Ann. Math. {\bf 160}, 433 (2004).
\bibitem{Amit} D.J. Amit, G. Parisi, and L. Peliti, Phys. Rev. B {\bf 27}, 1635 (1983).
\bibitem{Freund} H. Freund and P. Grassberger, Physica A {\bf 192}, 465 (1993).
\bibitem{New-Bark} M.E.J. Newman and G.T. Barkema {\it Monte Carlo Methods in Statistical Physics} 
  (Oxford University Press, New York 1999).
\bibitem{Ziff} R.M. Ziff, Computers in Physics, {\bf 12}, 385 (1998).
\bibitem{Belius-a} D. Belius, Probab. Theory \& Related Fields {\bf 157}, 635 (2013).
\bibitem{Belius-b} D. Belius and N. Kistler, Probab. Theory \& Related Fields {\bf 167}, 1 (2014).
\bibitem{Dvoretsky} A. Dvoretsky and P. Erd\"os, in {\it Proc. of the $2^{nd}$ Berkeley 
   Symposium of Mathematical Statistics and Probability}, p.353 (Princeton University Press, 
   Priceton N.J. 1951).


\end{thebibliography}
\end{document}